\documentclass{aa}
\usepackage[varg]{txfonts}
\usepackage{natbib}
\bibpunct{(}{)}{;}{a}{}{,}

\usepackage{booktabs}
\usepackage{hyperref}
\usepackage{algorithm}
\usepackage{algorithmic}

\title{Kepler-404: A near-resonant pair of sub-Neptunes}
\subtitle{Reinterpreting a false-positive signal with the novel detection method RIVERS.radon}

\author
{
	Yannick Eyholzer\inst{1},
	Marylyn Rosenqvist \inst{1} ,
	Adrien Leleu\inst{1},
	Vitaliy Kinakh \inst{2} and
	Slava Voloshynovskiy \inst{2}
}	

\authorrunning{Y. Eyholzer et al.}
\titlerunning{The near-resonant pair of Kepler-404 characterised with RIVERS.radon}

\institute
{
	Observatoire de Genève, Université de Genève, Chemin Pegasi 51, 1290 Versoix,
	Switzerland \and Campus Battelle, Université de Genève,  Route de Drize 7, 1227 Carouge, Switzerland}

\date{Received ?/ Accepted ?}

\abstract 
{According to the current understanding of planet formation and evolution, (near)-resonant systems should be a common occurrence in nature. However, pairs of planets with period ratios close to commensurability are rarely observed. In this paper, we discuss how observational bias caused by transit timing variations (TTVs) can affect the detection of low signal-to-noise ratio (S/N) planets in transit surveys, playing a role in the lack of representation of these systems. We present an alternative detection method, RIVERS.radon, which overcomes the effect of TTVs on planet detectability by not relying on an exact periodic solution for the transit signal. It leverages 2D representations of the light curve and approximates the transits with a set of local periodic solutions with the use of 2D cross-correlation. We present an application of RIVERS.radon to the near-resonant system Kepler-404, showing that the signal of the false-positive candidate KOI-2768.03 belongs to the planet Kepler-404b and that it is in a near 5:4 resonance with its companion Kepler-404c, causing TTVs of amplitudes $117.64$ and $110.09$ minutes, respectively. With a photodynamical fit, we retrieve a pair of similar planets, respectively of radii $1.68_{-0.09}^{+0.09}$ and $1.69_{-0.10}^{+0.10} R_{\oplus}$ and of masses $1.91_{-0.33}^{+0.41}$ and $2.34_{-0.40}^{+0.47} M_{\oplus}$. We find that these planets exhibit typical (near)-resonant features: their densities are lower than the bulk population of exoplanets, and their orbits are nearly coplanar. Interestingly, the planets are found outside the resonance but at a significant distance from the equilibrium point of the $(X_{\mathrm{res}}, \delta)$ plane of the Hamiltonian, which is consistent with post-disk instability but somewhat at odds with the observed coplanarity.}

\keywords{planets and satellites: detection --
	planets and satellites: dynamical evolution and stability --
	planets and satellites: fundamental parameters --
	methods: data analysis --
	techniques: photometric --
	stars: individual: Kepler-404}

\begin{document}
	
	\maketitle 
	\nolinenumbers
	\section{Introduction}

	\label{sect:intro}
	
	The transit method has proven to be the most successful method for the discovery of exoplanets so far. By continuously monitoring the stellar flux over time, it is possible to detect periodic dips in the brightness of the star, symptomatic of a planet transiting across the star along the observer's line of sight.  When these dips in flux are caused by a planet, it is possible to assess the relative size of the planet to the host star. Given that planets follow Keplerian orbits, the elapsed time between two successive transit events is constant and can be measured to infer the orbital period.
	
	When a single transit signal is too faint relative to the noise, i.e. the signal-to-noise ratio (S/N) is too low, the periodic nature of the signal can be leveraged to enhance the S/N significantly. A notable example of this approach is the Box-fitting Least Squares (BLS) algorithm \citep{kovacs_box-fitting_2002}, which improves the detection of low-S/N single transit events by folding the light curve over a grid of trial periods, epochs, and durations and identifying periodic signals by effectively stacking all transits.
	
	However, in multi-planetary systems, the assumption of strictly Keplerian orbits no longer holds, as mutual gravitational interactions between the planets induce perturbations \citep{agol_detecting_2005}. The magnitude of these perturbations depends on the planetary masses and the system's configuration, and is significantly amplified for (near)-resonant systems as the perturbative effects accumulate over time \citep{lithwick_extracting_2012,nesvorny_dynamics_2016, deck_transit_2016}. This leads to quasi-periodic orbits, deviating from the Keplerian periodicity. Observationally, these deviations manifest as transit timing variations (TTVs). 
	
	Standard algorithms relying on the folding of the light curves at a fixed period are fundamentally limited when applied to systems exhibiting TTVs and often introduce biases. This can result in smeared transit signals \citep{garcia-melendo_potential_2011}, leading to inaccurate determination of the planetary parameters, or, in other cases, a complete non-detection of the planetary signals or mistaking the signal for a false positive \citep{leleu_alleviating_2021, leleu_alleviating_2022}. \cite{kane_visual_2019} discuss in their work how the under-representation of planets with small radii and large TTVs in the Kepler dataset is consistent with an observational bias. The smearing effect is expected to be strongest when the TTVs have a periodicity comparable to or smaller than the observational baseline timescale, with an amplitude similar to or above the transit duration. Thus, this effect will be particularly strong for small (near)-resonant planets in transit surveys, as the transits of these planets are too weak to be detectable individually, possibly resulting in observational incompleteness.

	This is particularly problematic as resonant systems are of strong scientific interest. Theoretical models and simulations suggest that resonance chain capture is a common occurrence during the formation of planetary systems \citep{lee_dynamics_2002, terquem_migration_2007}.
	Observationally, we see that there is a pile-up of planet pairs near first-order mean motion resonances (MMRs), but systems that are exactly in resonance are less common \citep{fabrycky_architecture_2014}. Instabilities in chains are capable of breaking the resonance state of a system \citep{mills_resonant_2016, izidoro_breaking_2017} and can explain the lack of representation of these resonant systems in our current population of exoplanetary systems. However, observational effects could also play an important role.  
	
	Resonant systems are able to provide unique insights into the formation and evolution history of planetary systems, granting better constraints on current models. Detecting planets with TTVs makes it possible to identify these systems and increase their representation. Additionally, as the gravitational interactions between the planets are encoded in the TTVs, it is possible to retrieve the mass of the interacting planets in some cases. This proves to be incredibly powerful as full characterisation of the planet's mass and radius, thus density, is possible with a single method, while it would require the combined usage of multiple techniques to allow this in normal cases. Moreover, the dynamical signature of an interacting planet on another, transiting one makes it possible to infer its presence even if the geometry of its orbit prevents it from transiting. TTVs are indeed an additional rich source of information.
	
	To address these challenges and leverage current instruments and available data (Kepler, TESS) and to be ready for coming opportunities (e.g. PLATO), it is important to develop detection methods that  relax the strict assumption of periodic transits to mitigate systematic bias arising from periodic folding when applied to resonant systems. With this aim, several methods and algorithms have been developed \citep{carter_quasiperiodic_2013, leleu_alleviating_2021, leleu_alleviating_2022}. These approaches, used alone or combined, can enhance the sensitivity to detect dynamically active planets, allow accurate characterisation of the planetary system, and permit detailed dynamical studies through the initial measurement of the TTVs.
	
	This paper introduces river diagrams in Section \ref{sect:repesentation} and describes in Section \ref{sect:method} how to exploit them with a novel method. It then presents an application to a real system, Kepler-404, in Section \ref{sect:application}. Finally, a photodynamical analysis of the system is presented and discussed in Section \ref{sect:characterisation}.

	\section{River diagram: a change of representation}
	\label{sect:repesentation}
	As introduced in Section \ref{sect:intro}, standard methods for low-S/N transit detection rely on the assumed periodicity of the transits. These methods usually work by fitting a periodic model, or equivalently, folding the light curve on a chosen period and fitting a single transit model. This is efficient, as the model is reduced to 4 parameters for simple box-shaped transits, being the period, epoch, duration, and depth of the transit. The number of parameters increases for more sophisticated transit models, but the assumption on the periodicity of the signal remains. Under the assumption of a box-shaped transit signal, uniform scatter, and uncorrelated noise, the S/N of a single transit event is estimated with
	\begin{equation}
		S/N_i = \sqrt{N_{p}}\frac{D}{\sigma}
	\end{equation}
	with ${N_{p}}$ being the number of data points in the transit, ${D}$ the depth of the transit and ${\sigma}$ the standard deviation of the light curve. Folding the light curve on the exact period effectively increases the number of points in transit in the phase space, increasing the S/N. For ${N_t}$ transit events present in the light curve, the new S/N can be estimated with
	\begin{equation}
		\label{eq:snr_fold}
		S/N_{\mathrm{fold}}= \sqrt{N_t N_{p}} \frac{D}{\sigma} =\sqrt{N_t}  S/N_i
	\end{equation}
	which holds in the case of periodic transits within the cadence precision of the observation, as each transit perfectly aligns with the others when folded on the correct period. 
	
\begin{figure}
		\centering
		\includegraphics[width=.8\linewidth]{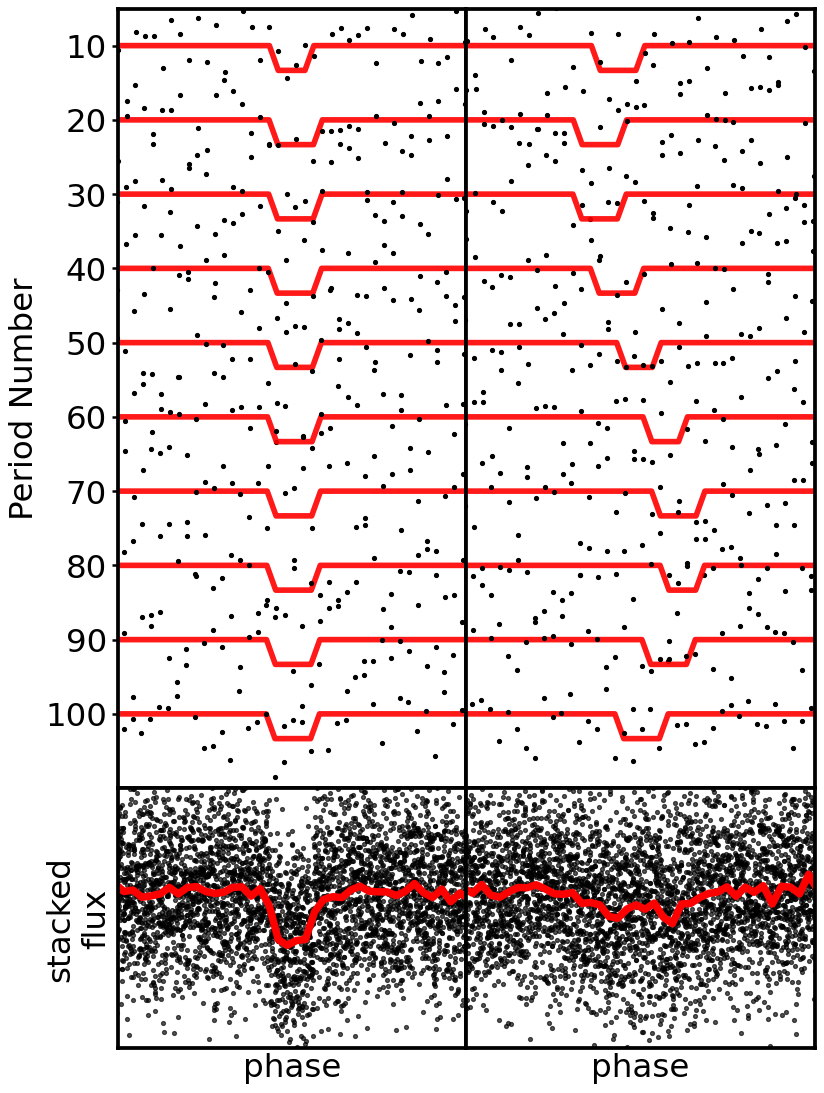}
		\caption{Stacked transit plots for the periodic case on the left and quasi-periodic case on the right. The bottom plots show the corresponding phase-folded light curves, with the binned data in red. The quasi-periodic case clearly displays the smearing effect.}
		\label{fig:smearing}
	\end{figure}

	However, in the quasi-periodic case, Equation \ref{eq:snr_fold} is no longer true. Indeed, there is not a single period that can express the time between all the transits, hence transits cannot be perfectly aligned, as portrayed in the top part of Figure \ref{fig:smearing}. Consequently, folding the light curve on any period will lead to a smearing effect in the phase-folded light curve, visible on the bottom part of Figure \ref{fig:smearing}. The resulting ``transit'' is less deep and wider \citep{garcia-melendo_potential_2011}. If the observation time encompasses a complete period of the perturbation, the reduced S/N in that case can be approximated \citep{leleu_alleviating_2021}:
	\begin{equation}
		\label{eq:snr_ttv}
		S/N_{\mathrm{TTV}}= \frac{1}{1 + T_{\mathrm{dur}}/\sigma_{\mathrm{TTV}}} S/N_{\mathrm{fold}}
	\end{equation}
	where ${T_{\mathrm{dur}}}$ is the duration of the transit and $\sigma_\mathrm{TTV}$ the amplitude of the TTVs. It shows that working with phase-folded light curves in the presence of TTVs is not efficient, as the resulting smearing leads to a wrong characterisation of the signal, or renders it undetectable, and additionally erases the dynamical signature of the planet interactions that could be exploited.

\begin{figure}
		\centering
		\includegraphics[width=.8\linewidth]{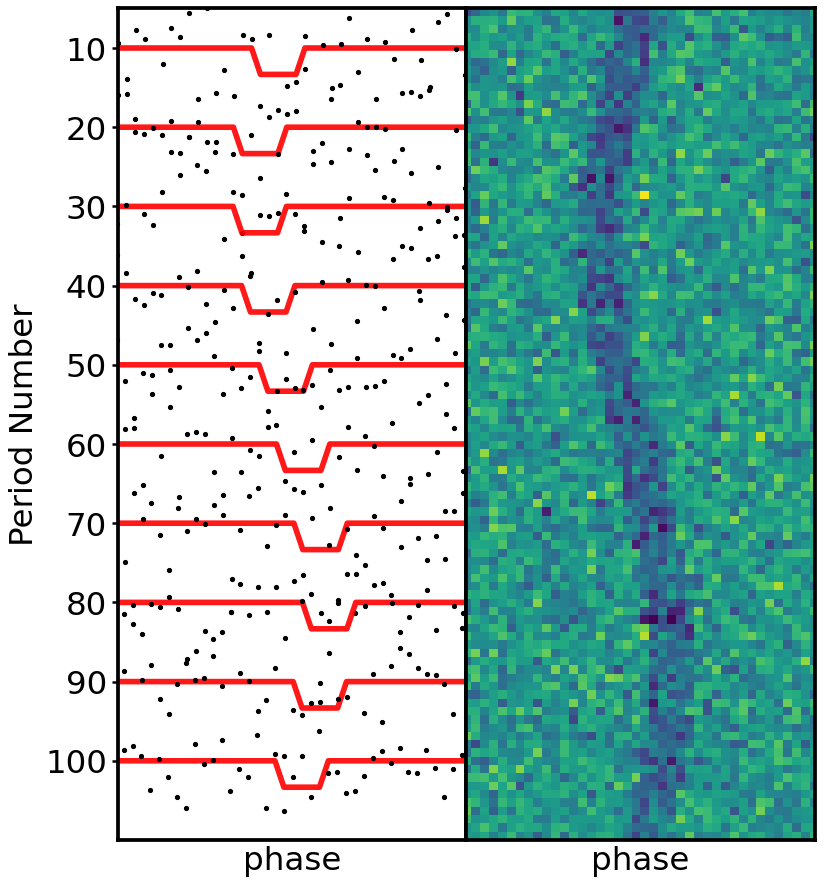}
		\caption{Stacked phase of the light curves with the corresponding river diagram. The S/N in the river diagram example was increased to emphasise the transit tracks.}
		\label{fig:river_diagram}
	\end{figure}

	The proposed change of representation exploits the fact that TTVs are small compared with the orbital period: an average period $P_\mathrm{mean}$ can be found and subtracted from the transit times, so that only the perturbative component remains.
	By representing the light curve in a 2D form, as shown in the top part of Figure \ref{fig:smearing}, with a folding period close to $P_{\mathrm{mean}}$, the transits become vertically coherent,  as visible in Figure \ref{fig:river_diagram}. The right representation is the same image as the left, except that the flux is represented in pixel intensity. This representation is known as river plots or river diagrams \citep{carter_kepler-36_2012}. If the river diagram is created with a period that is close to $P_{\mathrm{mean}}$, each transit will be in phase and close enough from one row to another to be shown as a vertical feature in the image. Identifying these transit tracks would be equivalent to extracting the individual transit times.
	
	\section{RIVERS.radon}
	\label{sect:method}
	RIVERS.radon is a method inspired by the general Radon transform \citep{radon_determination_1986}. The original Radon transform works by integrating the image intensity along straight lines over the whole image, parameterised by an angle $\theta$ and offset $\rho$, mapping the image into a ($\rho, \theta$) feature space. The transform enables the extraction of straight line features in images, and is especially suitable for feature extraction in noisy images \citep[e.g.][]{nir_optimal_2018}. \cite{copeland_localized_1994} discuss how features that are significantly shorter than the image dimension, or span the entire image but that have some curvature, can be hard to detect with this method. Additionally, the global projection erases local information such as the feature edges along the projection line. Finally, images with high levels of noise will amplify these problems. Thus, they adapted the Radon transform to be able to detect local features in images, by limiting the projection length, and adding a new parameter in the transform to preserve the positional information of the feature in the image.
	
	This work presents an implementation of the localised Radon transform to work on river diagrams with the aim of identifying transit tracks and extracting individual transit times. Transit tracks span the entire river diagram on the vertical axis when the image is created on a period close to that of the planet. However, the TTVs translate to curvature in the transit tracks, hence the need for a localised approach, as the features are indeed not straight lines, but can still be well approximated locally with a set of shorter line segments. In other words, we assume that we can model successive transit times with a local periodic solution. The RIVERS.radon method relies on a bank of short oriented line kernels $\{K^\ell\}$ of fixed length and width. The local projection of the image intensity along a line segment of a given orientation is measured with cross-correlation of the kernels with the river diagrams. A high correlation response indicates that a line of specific orientation at a given location of the river diagrams matches a feature, potentially being a transit track. A short summary of the method can be found in Algorithm \ref{alg:radon}, and a more detailed explanation in the following subsections.
	
	The algorithm is publicly available\footnote{\url{https://github.com/yannholzer/RIVERS.radon.git}}.
	
	\begin{algorithm}
		\caption{RIVERS.radon transit-track extraction}
		\label{alg:radon}
		\begin{algorithmic}[1]
			\REQUIRE river diagram $I \in \mathbb{R}^{H\times W}$, created from detrended, normalised light curve at folding period $P_{\mathrm{mean}}$; kernel length and width;
			number of orientations $L$ and angular range $\pm\theta_{\max}$;
			dilation window $d$; erosion window $e$ (with $d<e$);
			target signal-to-noise $\mathrm{S/N}_{\mathrm{target}}$.

			\STATE Build oriented binary line kernels $\{K^{(\ell)}\}_{\ell=1}^{L}$.
			\STATE Pad $I$ with helical (phase-wrapped) boundary conditions (Eq.~\ref{eq:pad}).
			\FOR{each orientation $\ell = 1,\dots,L$}
			\STATE Compute local correlation $C^{(\ell)} = \tilde{I}\star K^{(\ell)}$ and average $A^{(\ell)} = C^{(\ell)}/n_\ell$.
			\STATE Estimate local depth $T^{(\ell)} = 1 - A^{(\ell)}$.
			\STATE Threshold to binary map $B^{(\ell)}$ using $\tau_\ell$ from $\mathrm{S/N}_{\mathrm{target}}$ (Eq.~\ref{eq:threshold}).
			\STATE Apply morphological cleaning: dilation ($d$) then erosion ($e$) $\rightarrow E^{(\ell)}$, using helical padding on each input.
			\ENDFOR
			\STATE Reconstruct confidence map $R = \sum_{\ell=1}^{L}(\tilde{E}^{(\ell)}\star K^{(\ell)})$.
			\STATE Convert $R$ to a confidence time series and extract the transit-time sequence $\{t_i\}_{i=1}^{M}$ (Appendix).
		\end{algorithmic}
	\end{algorithm}
	
	\subsection{Problem formulation}
	Let the river diagram created from a detrended and normalised light curve be represented as a 2D image $I \in \mathbb{R}^{H \times W}$, $H$ being the number of periods (rows) and $W$ the number of phase bins (columns). Each pixel $I(p,q)$ corresponds to the observed flux at the period index $p$ in the phase bin index $q$.
	
	In this representation, for a folding period of the river diagram close enough to that of the planet, the successive transits of a planet form an approximately coherent track in the form of a sequence of pixels with lower-than-average intensity across rows. In the presence of TTVs, this track is generally not perfectly straight but remains locally well approximated by short line segments.
	
	The objective of this method is therefore to detect the transit tracks in the image and to convert them into a sequence of estimated individual transit times.
	
\begin{figure}
		\centering
		\includegraphics[width=.8\linewidth]{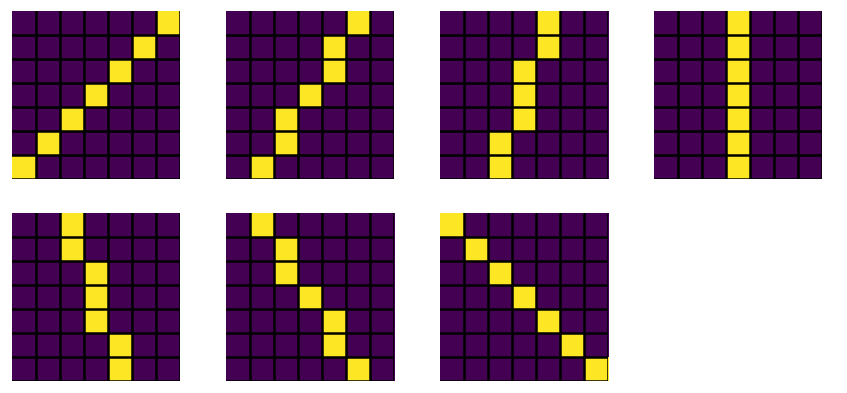}
		\caption{Example of a set of kernels used by RIVERS.radon.}
		\label{fig:kernels}
	\end{figure}

	To this end, we introduce a family of oriented binary kernels:
	\begin{equation}
		K^{(\ell)} \in \{0,1\}^{M \times N}, \quad \ell = 1,\dots,L,
	\end{equation}
	where each kernel corresponds to a line segment of fixed length and width and of a given orientation, as shown in Figure \ref{fig:kernels}. Kernels of longer line lengths will be more efficient at retrieving signals from the ambient noise. However, they will be less sensitive to the local variations of the transit tracks caused by the TTVs. On the other hand, very short lines will precisely capture variations of the features, but are also more sensitive to false positives caused by local variations in the river diagram. The kernel width should be adapted to the ratio between the binning used to create the river diagram and the expected transit duration. If a transit track is expected to span multiple pixels of the river diagrams, the width of the kernel can be increased to enhance the cross-correlation response. Otherwise, the width should stay minimal so as not to mix the transit pixels with the noise, reducing the response. The number of orientations should be sufficient to uniformly span the chosen angular range. Finally, the angular range should be adapted to the expected maximum orientation of the transit tracks, which depends firstly on the precision of the folding period of the river diagram relative to the actual period of the planet, but also on the expected TTV amplitude of the transit tracks.

	Let
	\begin{equation}
		n_\ell = \sum_{u=0}^{M-1} \sum_{v=0}^{N-1} K^{(\ell)}(u,v)
	\end{equation}
	be the number of active elements in the kernel $K^{(\ell)}$. A set of kernels is created to have an ensemble of lines going from plus or minus a chosen angle relative to the vertical axis. The kernels are constructed to be vertically symmetric. The width of the line can be changed but is also required to be symmetrical horizontally.

	Hence, kernels always have odd dimensions. This is an important condition as multiple operations used with this method do not preserve the image dimensions. For example, the output dimensions of the cross-correlation operation are:
	\begin{align*}
		H' &= \left[\frac{H + 2P_H - M}{S}\right] + 1 \\
		W' &= \left[\frac{W + 2P_W - N}{S}\right] + 1 
	\end{align*}
	where $P$ is the padding of the image in each dimension, $S$ the stride of the cross-correlation kernels and $(M,N)$ the dimensions of the kernel. For $S=1$, we see that $H'=H$ and $W'=W$ if \begin{align*}
		P_H &= \lfloor M / 2 \rfloor\\
		P_W &=\lfloor N / 2 \rfloor 
	\end{align*}
	where $\lfloor .. \rfloor$ denotes the floor operation. The equality above is true only if $M,N$ are odd. We define the padding function Pad:  $\mathbb{R}^{H \times W}  \to \mathbb{R}^{(H+2\lfloor h/2 \rfloor) \times (W+2\lfloor w/2 \rfloor)}$ such that:
	\newcommand{\floor}[1]{\lfloor #1 \rfloor}
	\begin{equation}
		\label{eq:pad}
		\text{Pad}(I,\, h,\, w,\, c)(p, q) = 
		\begin{cases}
			I\!\left(p', q' \bmod W \right) & \text{if } p' \in [0, H) \\[6pt]
			c & \text{otherwise}
		\end{cases}
	\end{equation}
	where 
	\begin{equation*}
		p' = p - \lfloor h/2 \rfloor + \lfloor q'/W \rfloor, \quad q' = q - \lfloor w/2 \rfloor
	\end{equation*}
	This construction induces a helical topology, ensuring the continuity of transit tracks across phase boundaries. Indeed, a signal going out of bounds on the right side of the river diagram will start again on the next row below. Conversely, a signal starting on the left side originates from the previous row above. Thus, the helical padding ensures this logic is respected and allows a feature on the edges of the image to still have a strong cross-correlation response.
	In contrast, the constant padding ensures the response stays minimal out of bounds of the image. In practice, padding with the continuum value $c=1$ will cause the estimated depth (Equation \ref{eq:estimated_depth}) to vanish in these areas. The padding function preserves the image dimension.

	\subsection{Local Radon transform}

	For each kernel $K^{(\ell)}$, we compute a local correlation response over the padded river diagram. We define the discrete correlation operation:
	\begin{equation}
		(f \star g)(p,q) = \sum_{u=0}^{M-1} \sum_{v=0}^{N-1} f(p+u, q+v)g(u,v)
	\end{equation}
	Using this definition the response map is given by
	\begin{equation}
		C^{(\ell)}(p,q) = (\widetilde{I} \star K^{(\ell)})(p,q)
	\end{equation}
	where $\widetilde{I}= \text{Pad}(I, M, N, 1)$. This quantity sums the pixels of the river diagram $I$, i.e. the flux of the light curve,  that match the position of a kernel centred on the position $(p,q)$ for an orientation $\ell$. To obtain the average flux, we divide by the number of active kernel elements:
	\begin{equation}
		A^{(\ell)}(p,q)
		=
		\frac{C^{(\ell)}(p,q)}{n_\ell}.
	\end{equation}
	Since transits correspond to dips in flux, the transit tracks translate to lower-than-average structures in the image. Thus we can estimate the local depth by taking the difference from one
	\begin{equation}
		\label{eq:estimated_depth}
		\mathcal{T}^{(\ell)}(p,q) = 1 - A^{(\ell)}(p,q),
	\end{equation}
	assuming the river diagrams are created from normalised flux. Collecting all orientations yields a cube of estimated depths.
	\begin{equation}
		\mathcal{T} \in \mathbb{R}^{L \times H \times W}.
	\end{equation}
	
	\subsection{Thresholding}
	The detection scores are thresholded into binary detections:
	\begin{equation}
		B^{(\ell)}(p,q)
		=
		\begin{cases}
			1, & \mathcal{T}^{(\ell)}(p,q) \ge \tau_\ell, \\
			0, & \text{otherwise}.
		\end{cases}
	\end{equation}
	where $\tau_\ell$ is a target depth. As using a target S/N is more intuitive, we can estimate $\tau_\ell$ from a chosen $\mathrm{S/N}_{\mathrm{target}}$. Recalling Equation \ref{eq:snr_fold}, we can isolate the expected transit depth:
	\begin{align}
		\mathrm{S/N}_{\mathrm{target}} &= \frac{\tau_\ell}{\sigma_I} \sqrt{n_\ell} \\\Rightarrow \tau_\ell &=
		\label{eq:threshold}
		\frac{\mathrm{S/N}_{\mathrm{target}} \sigma_I}{\sqrt{n_\ell}}
	\end{align}
	where $\sigma_I$ is the standard deviation of the river diagram $I$. The target S/N is computed for an assumed number of transiting points, given by the number of activated pixels $n_\ell$ in a kernel of a specific orientation. It can be understood as a target averaged cross-correlation response threshold, normalised by the standard deviation of the input image. Thus, smaller kernels will have a higher false-positive rate than bigger kernels since a smaller number of pixels of the image is more likely to have an averaged cross-correlation response that passes the threshold due to the noise. In contrast, bigger kernels will have a lower false-positive rate, at the cost of a reduced sensitivity to the curvature of the signals.
	
	\subsection{Morphological cleaning}
	The binary maps $\{B^{(\ell)}\}_{\ell=1}^{L}$ may still contain isolated activations caused by local fluctuations or imperfect alignments. To improve spatial coherence, we apply a simple morphological cleaning step independently on each orientation channel.
	
	First, we define the dilation maps $\{D^{(\ell)}\}_{\ell=1}^{L}$ of the padded binary maps $\{\widetilde{B}^{(\ell)}\}_{\ell=1}^{L}$. For a chosen dilation window $d$, each binary map is padded with $\widetilde{B}^{(\ell)}=\text{Pad}(B^{(\ell)}, 0, 2d, 0)$ and the elements of the dilation maps are defined by the operation
	\begin{equation}
		D^{(\ell)}(p, q) = \max_{s \in[-d, d]} \widetilde{B}^{(\ell)}(p, q+s)
	\end{equation}
	which returns 1 if any element in the window is non-zero, effectively connecting nearby elements of $\{B^{(\ell)}\}_{\ell=1}^{L}$ along the phase direction and compensating for small discretisation errors or local interruptions of the track.
	
\begin{figure}
		\centering
		\includegraphics[width=.8\linewidth]{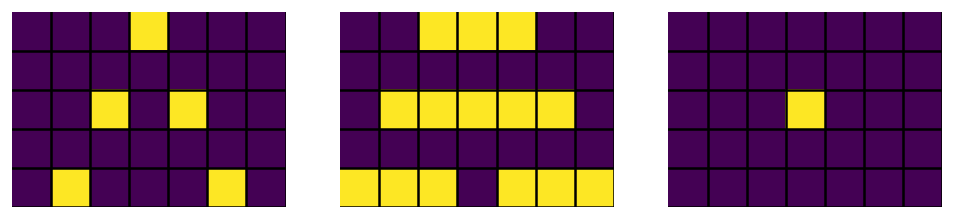}
		\caption{Example of the cleaning operation. \textit{left:} initial condition. \textit{middle:} after the dilation operation with $d=1$. \textit{right:} after the erosion operation with $e=2$.}
		\label{fig:cleaning_operation}
	\end{figure}

	Second, we define the erosion maps $\{E^{(\ell)}\}_{\ell=1}^{L}$ of the padded dilation maps $\{\widetilde{D}^{(\ell)}\}_{\ell=1}^{L}$.  For a chosen erosion window $e$, each dilation map is padded with $\widetilde{D}^{(\ell)}=\text{Pad}(D^{(\ell)}, 0, 2e, 0)$, the elements of the erosion maps are computed as
	\begin{equation}
		E^{(\ell)}(p,q) = \min_{s \in[-e, e]} \widetilde{D}^{(\ell)}(p, q+s) 
	\end{equation}
	which returns 1 only if all pixels in the horizontal window are activated, effectively keeping local regions of the binary maps that are fully activated and discarding the others.  A simple example of the morphological cleaning is visible in Figure \ref{fig:cleaning_operation}. The cleaning is effective as long as $d < e$.

	\subsection{Reconstruction}
	The cleaned padded binary detections $\{\widetilde{E}^{(\ell)}\}_{\ell=1}^{L}$, where the elements are created as $\widetilde{E}^{(\ell)}=\text{Pad}(E^{(\ell)}, M, N, 0)$, are projected back into image space. This is achieved by taking the cross-correlation of each binary map with its corresponding kernel and summing across orientations
	\begin{equation}
		R(p,q) = \sum_{\ell=1}^{L} (\widetilde{E}^{(\ell)} \star K^{(\ell)})(p,q).
	\end{equation}
	
	The reconstructed image $R \in \mathbb{R}^{H \times W}$ can be interpreted as a confidence map: large values indicate pixels that are repeatedly supported by multiple locally consistent line detections.
	
	\subsection{From mask to transit times}
	To extract transit times from the confidence map, we use the corresponding time matrix of the input river diagram and convert the confidence map into a confidence time series. Following that, we extract a sequence of estimated transit times. The first transit time is extracted from the maximum confidence pixel time location. The previous and next transit times are then extracted from the maximum confidence time in adjacent intervals, given by the folding period and a chosen tolerance window to account for the potential offset caused by the TTVs. If a new transit time is found, the next interval is computed from it; otherwise, it is computed from the last transit time found, stepping forward by one period. The process ends when both the start and end of the time series are reached. See Appendix \ref{sect:appendix} for a detailed explanation.
	
	It is important to note that these timings are proxies, and will be used as initial conditions for the following photodynamical fits. They do not carry formal uncertainties; however, the bin size used in the river diagram can be used as an inherited uncertainty.
	
	\section{Application of RIVERS.radon to the Kepler-404 light curve}
	\label{sect:application}
	\subsection{Overview of the system}
	At the time of writing, Kepler-404 hosts two confirmed transiting planets, Kepler-404b and Kepler-404c \citep{rowe_validation_2014}, and a false positive, KOI-2768.03, flagged ``Not Transit-Like'' by the Kepler pipeline. The stellar properties can be found in Table \ref{tab:stellar_properties}. 
	
	The candidates' status and periods are visible in Table \ref{tab:planets_periods}.  It is striking how close the period of Kepler-404b,  $P_\mathrm{b}=11.829851$ days, is to that of the false positive $P_\mathrm{fp}=11.83271013$ days. Additionally, looking at the period ratio of Kepler-404b and Kepler-404c, 
	\begin{equation}
		\frac{P_\mathrm{c}}{P_\mathrm{b}} = \frac{14.751166}{11.829851} = 1.2469 \approx \frac{5}{4} = 1.25
	\end{equation}
	we see the planets are close to a 5:4 first-order mean motion resonance.  These characteristics hint that the false positive may be worth investigating.

	\begin{table}[!h]
		\centering

		\renewcommand{\arraystretch}{1.5}
		\caption{Planetary candidates in the Kepler-404 system.}
		\label{tab:planets_periods}
		\begin{tabular}{lll}
			\toprule
			Name &Status & Period [days] \\
			\midrule
			Kepler-404b&   confirmed                               &  11.829851$\pm$0.000209                      \\ 
			Kepler-404c & confirmed                                &14.751166$\pm$0.000261                        \\ 
			KOI-2768.03 &  false positive                           &  11.832710$\pm$0.000053                 \\ 
			\bottomrule 
		\end{tabular}
		\tablefoot{Status and orbital periods from the NASA Exoplanet Archive, retrieved at the time of writing.}
	\end{table}
	
	\begin{table}[!h]
		\centering
		\setlength{\tabcolsep}{10pt} 
		\renewcommand{\arraystretch}{1.5} 
		\caption{Stellar properties of Kepler-404.}
		\label{tab:stellar_properties}
		\begin{tabular}{lll}
			\toprule
			\multicolumn{3}{c}{ Kepler-404 (KIC 9008737)} \\
			\midrule
			Parameter & Value & Ref \\ \hline
			m$_\mathrm{Kep}$ [mag]& $15.228$ & 1 \\
			m$_\mathrm{V}$ [mag]& $15.479 \pm 0.126$ & 1\\
			m$_\mathrm{J}$ [mag]& $13.908 \pm 0.027$  & 1\\
			m$_\mathrm{H}$ [mag]& $13.527\pm0.032 $ & 1\\
			\midrule
			T$_\mathrm{eff}$ [K]&  $5326\pm 60$ & 2\\
			\text{[Fe/H]} [dex] & $-0.032 \pm 0.040$ & 2\\
			L$_\star$ [L$_\odot$] &  $0.476^{+0.038}_{-0.036}$ & 3\\
			\midrule
			M$_\star$ [M$_\odot$] & $0.870 \pm 0.020$ & 2\\
			R$_\star$ [R$_\odot$] & $0.813 \pm 0.031$ & 2\\
			t$_\star$ [Gyr] & $4\pm4$ & 2\\
			$\log g$ [cm\,s$^{-2}]$ & $4.554^{+0.022}_{-0.035}$ & 3\\
			$\rho_\star$ [$\rho_\odot$] & $1.617\pm 0.188$ & 2\\
			\bottomrule
		\end{tabular}
	\tablebib{\raggedright (1)~NASA Exoplanet Archive (\url{https://exoplanetarchive.ipac.caltech.edu}); (2)~\citet{eggenberger_geneva_2008}; (3)~\citet{berger_gaia-kepler_2020}.}
	\end{table}
	\begin{figure*}[h]
		\centering
		\includegraphics[width=0.71\linewidth]{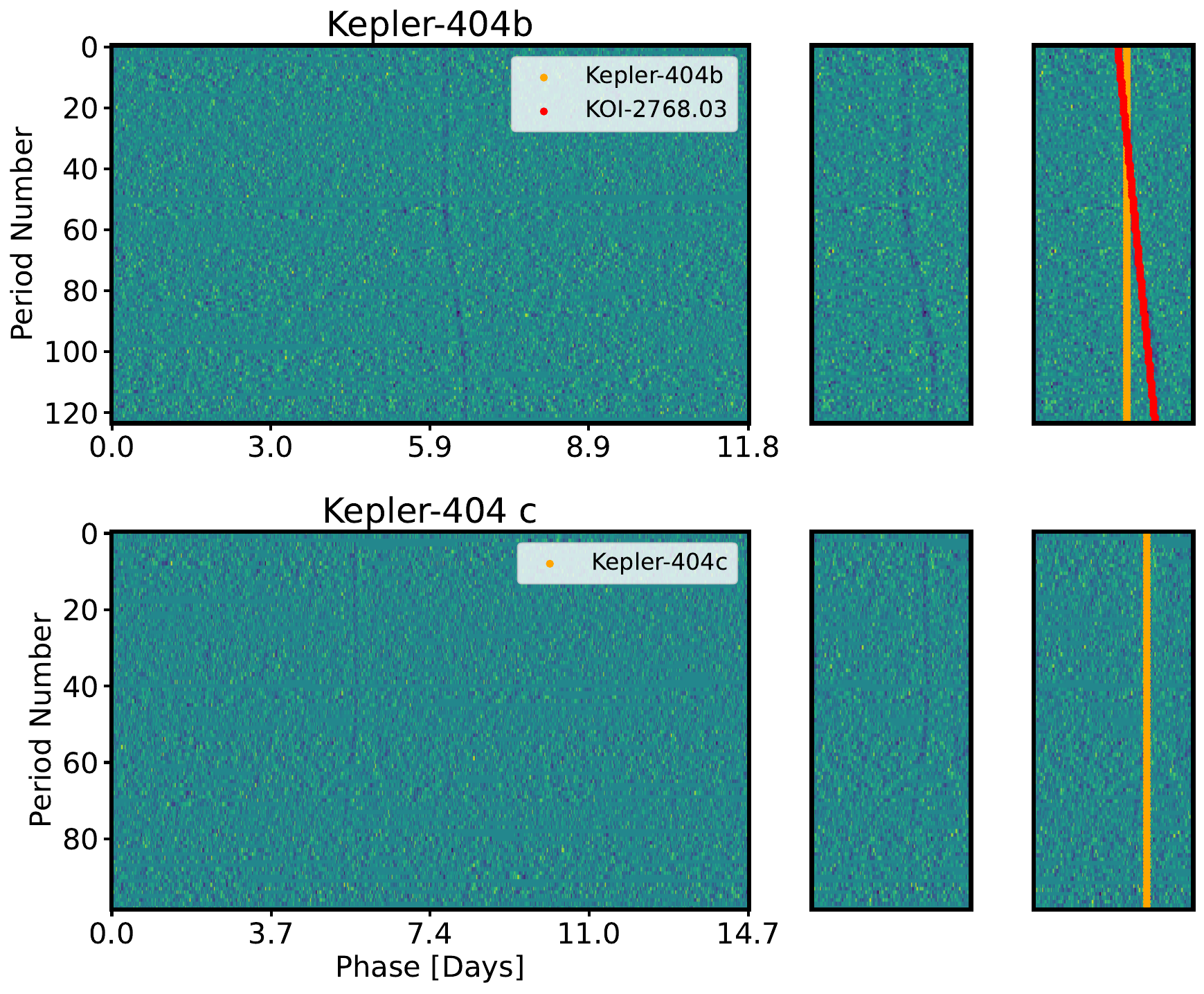}
		\caption{
			Kepler-404b and Kepler-404c river diagrams, in linear colour scale, respectively from top to bottom. The periods used to create the river diagram are those of the corresponding planet found in the archive, visible in Table \ref{tab:planets_periods}, and the bin size is 30 min. The left images are the full river diagrams, showing the phase in days on the horizontal axis and the period number on the vertical axis. The middle panels show a zoom on the transit tracks, and the right panels show the same zoom with the periodic solution found for the corresponding planet in orange and the ephemeris of the false positive KOI-2768.03 in red.
		}
		\label{fig:rd_kepler-404b_and_c}
	\end{figure*}
	
	\begin{figure*}[h]
		\centering
		\includegraphics[width=0.72\linewidth]{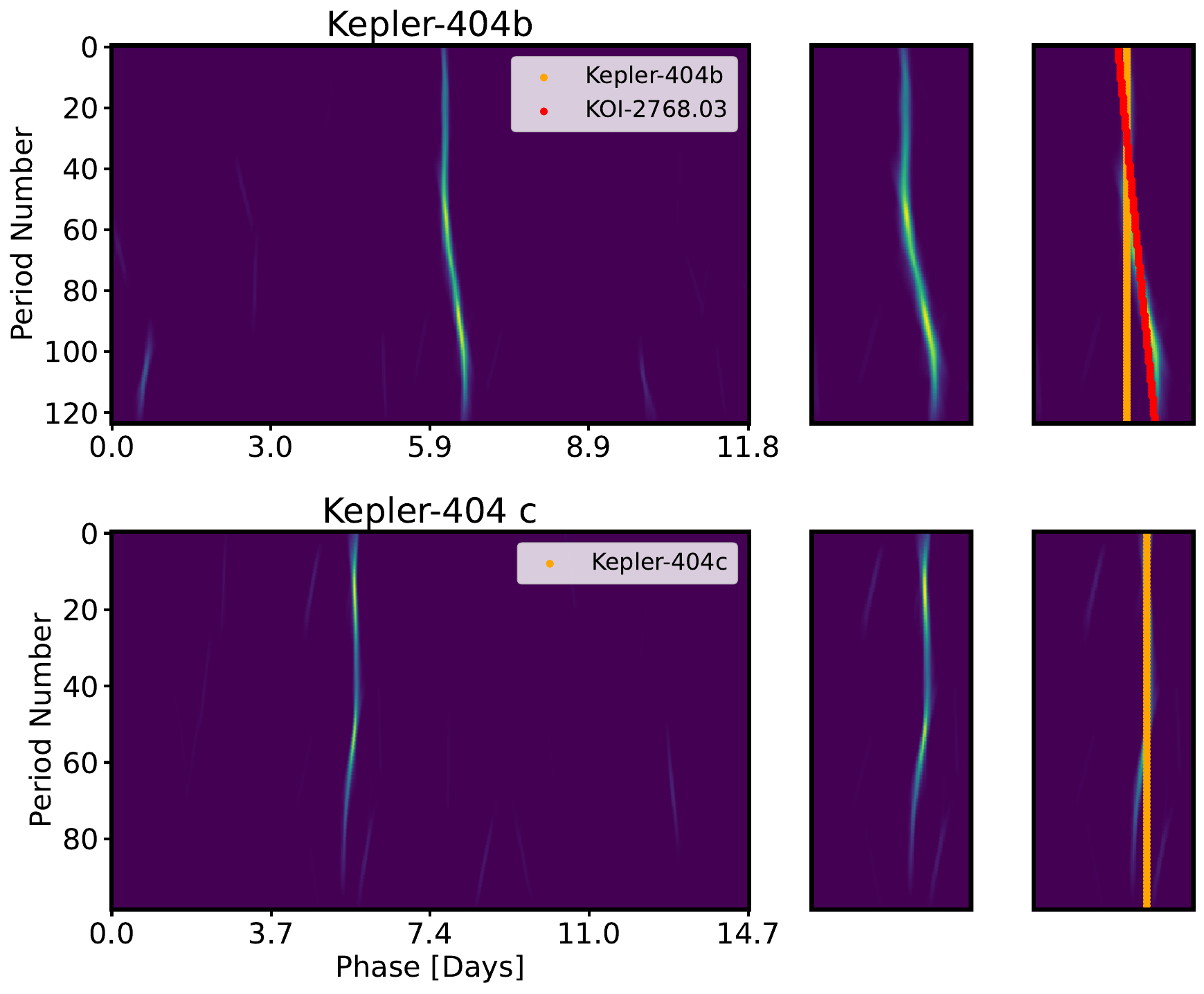}
		\caption{
			Output of the RIVERS.radon method using the river diagrams of Kepler-404b and Kepler-404c presented in Figure \ref{fig:rd_kepler-404b_and_c}.
		}
		\label{fig:reconstruct_kepler-404b_and_c}
	\end{figure*}

	\subsection{Preparation of the light curve}
	The long-cadence PDCSAP raw time series is downloaded using the lightkurve package. Copies of the light curve are created for Kepler-404b and Kepler-404c. For each light curve, we remove the signal of the other planet (including the false-positive signal of KOI-2768.03 when removing Kepler-404b) by propagating the ephemerides, calculating the transit times, and removing the time-series points within a window of 1.5 times the transit duration centred on the transit time.
	The processing of the light curve is done in chunks. In each quarter, we identify gaps greater than 2.5 hours, and split the quarter at these gaps into chunks, ensuring the detrending is the least affected by edge effects near the gaps. Each chunk is detrended with a spline method, with a detrending timescale of 3 times the estimated transit duration of the corresponding planet based on its period and the stellar density.
	In this way, trends on timescales longer than the b-spline timescale will be removed, while leaving the transits untouched. The detrending follows an iterative process to mask potential outliers beyond 4 standard deviations of the median of the light curve.
	All chunks are then stitched together into a single light curve. Finally, river diagrams are created for each planet using its corresponding light curve and the archive period of the planet. The river diagrams are created with a bin size of 30 min, slightly above the 29.4 min long cadence of the Kepler light curve to ensure at least one point per bin. The empty pixels in the river diagrams (caused by gaps and missing quarters) are filled with ones, ensuring a low cross-correlation response.
	
	Figure \ref{fig:rd_kepler-404b_and_c} shows the river diagrams of Kepler-404b and Kepler-404c, respectively. The transit tracks for both planets are visible thanks to the sufficiently high S/Ns of the transits. TTVs are noticeable in both diagrams, as the transit tracks are not straight lines but are curved. Inspecting the diagrams shows that the false-positive detection actually belongs to the same planetary signal as Kepler-404b. Indeed, due to the TTVs, one part of the signal was fitted with a periodic solution belonging to Kepler-404b, and another slightly different period retrieved another part of the signal. Finally, the transit tracks' variations are in anti-phase, which is expected for pairs of planets in first-order resonance \citep{lithwick_extracting_2012}.

	\subsection{Applying RIVERS.radon}
	Following the process explained in Section \ref{sect:method}, we forward the river diagram presented in Figure \ref{fig:rd_kepler-404b_and_c} to the RIVERS.radon method. A minimal set of RIVERS.radon parameters was chosen to assess the presence and shape of the TTVs: a line length equal to the river-diagram height, 1 pixel wide, no morphological cleaning ($d=e=0$) and $\mathrm{S/N}_{\mathrm{target}}=3$, 101 orientations spanning $\pm\frac{\pi}{3}$. The output of the method already displayed TTVs for both planets. Then, the kernel line width was increased to 3 pixels, adding morphological cleaning with $d=1$ and $e=2$ to reduce the false positives, and the orientation range was set to $\pm\frac{\pi}{4}$ as the TTV track curvature does not exceed this value. The line length was incrementally reduced together with the $\mathrm{S/N}_{\mathrm{target}}$ to visually constrain the transit tracks as tightly as possible without flooding the signals with false positives.
	The final set of parameters for both river diagrams is: line length of 25 pixels, line width of 3 pixels, $\mathrm{S/N}_{\mathrm{target}}=2$, 101 orientations spanning $\pm\frac{\pi}{4}$, dilation of $d=1$, erosion of $e=2$. The output of the method with the final set of parameters is shown in Figure \ref{fig:reconstruct_kepler-404b_and_c}.

\begin{figure}[!h]
		\centering
		\includegraphics[width=0.99\linewidth]{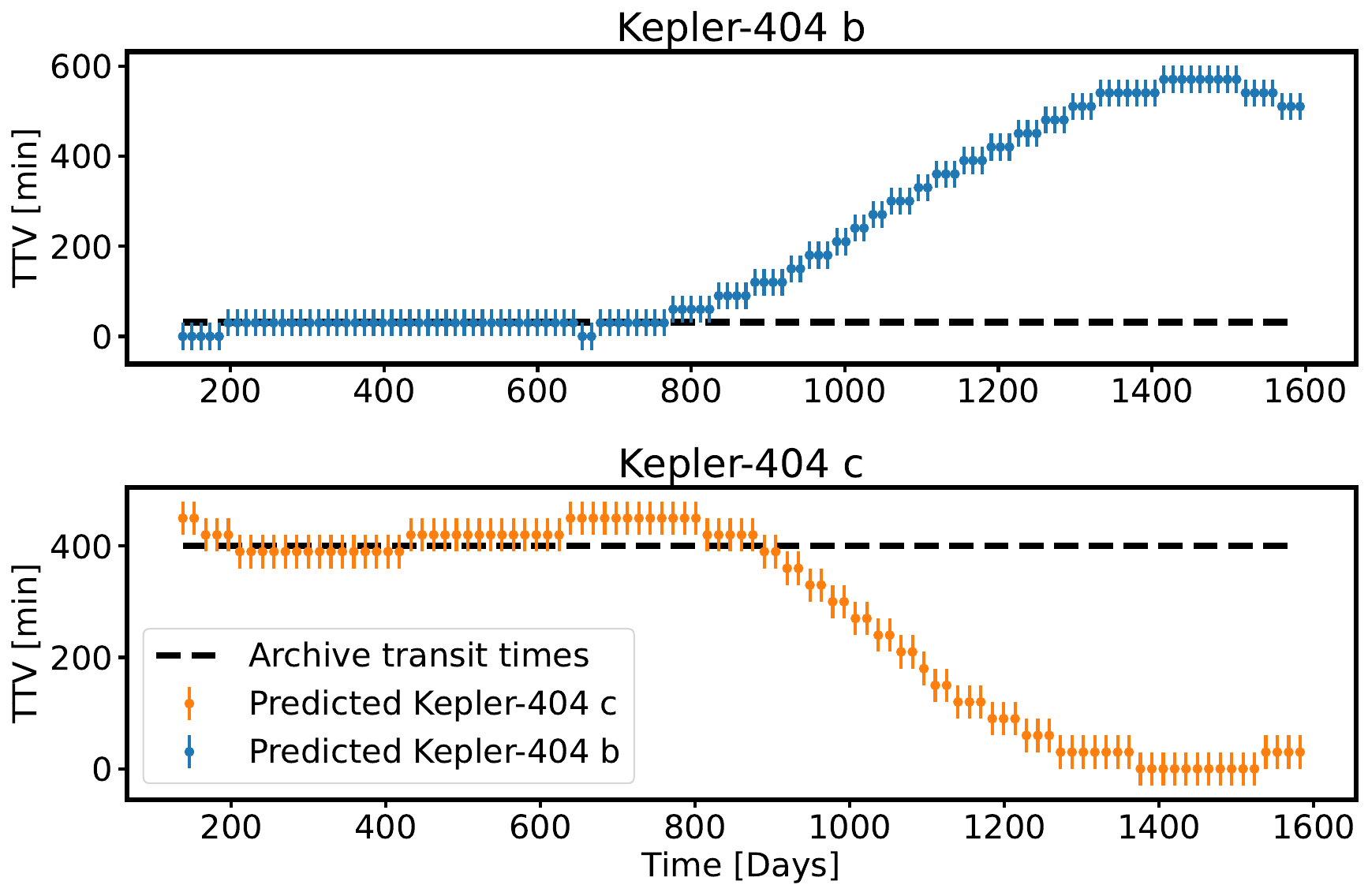}
		\caption{
			Deviation of the predicted transit times from those generated from the ephemerides in the archive for Kepler-404b and Kepler-404c.
		}
		\label{fig:ttv_archive}
	\end{figure}

	The deviation from the archive solutions is visible in Figure \ref{fig:ttv_archive}.
	
	\section{Characterisation of the planets}
	\label{sect:characterisation}
	\subsection{Fitting process}
	\label{sect:analysis}
	The characterisation of the planets is done in two steps following the process described in Section 2.5 of \cite{leleu_removing_2023}. First, we fit the proxy transit times retrieved by the RIVERS.radon method to gather orbital configurations and masses of both planets as approximate initial conditions that are used in a subsequent photodynamical fit. Both processes use the samsam algorithm \cite[see][]{delisle_harps_2018}\footnote{\url{https://gitlab.unige.ch/delisle/samsam}} to compute the posteriors and TTVFast \citep{deck_ttvfast_2014} to generate the transit times of a given orbital configuration. For the photodynamical fit, each sampled orbital configuration, including masses, is used to generate a light curve model that is fitted to the Kepler-404 raw light curve. The generated transit times from the sample configuration are used to locate the mid-transit times in the modelled light curve where the transit of each planet is injected using the batman package \citep{kreidberg_batman_2015}. Short-cadence data were used when available, and a super-sampling parameter set to 29.42 min was used for the long cadences. To prevent the fit from being affected by features introduced by detrending, the light-curve model is detrended together with the raw light curve by marginalising over the linear parameters (see \cite{leleu_removing_2023} for details)\footnote{\url{https://www.astro.unige.ch/~delisle/linmarg/doc/}}.
	
	\subsection{Robustness criterion}
	\label{sect:priors}
	Characterising robust masses and eccentricities with the TTV method via photodynamical fits can, in some cases, be hampered by a mass-eccentricity degeneracy, especially for near first-order resonances \citep{boue_degeneracy_2012, lithwick_extracting_2012}. These degeneracies are not straightforward to notice, and different priors can lead to seemingly good convergence of the sampling. This is a known problem that has been investigated, e.g. by \cite{hadden_kepler_2017, leleu_photo-dynamical_2024} and Rosenqvist et al. (in prep.). Criteria can be used to verify the robustness of the fit, i.e. the dependence of the parameter posterior distributions on the priors. Posterior distributions that stay close to one another while being generated from different priors are a good sign that the degeneracy is mitigated.
	In this work, we propose to compare three different priors, namely default, high eccentricity, and high mass, summarised in Table \ref{tab:priors}. The high-mass prior uses a uniform distribution to sample higher masses, while the high-eccentricity prior applies the same logic on the prior distribution of the eccentricities. The default samples are the ones that will be used for the rest of the discussion, using the eccentricity $\beta$ distribution introduced in \cite{kipping_parametrizing_2013}.
	\cite{hadden_kepler_2017} proposed considering a mass estimate robust if the median of the posterior coming from the high-mass prior lies within the 0.16--0.84 quantiles of the posterior using the default prior. \cite{leleu_removing_2023} followed with a criterion $\Delta_M = \max(\Delta_{M_{+}}, \Delta_{M_{-}})$, defined as:
	\begin{align*}
		\Delta_{M_{+}} &= \frac{m_{\mathrm{high},0} - m_{\mathrm{default}, 0}}{m_{\mathrm{default},+\sigma} - m_{\mathrm{default}, 0}} \\ \\
		\Delta_{M_{-}} &= \frac{m_{\mathrm{default},0} - m_{\mathrm{low}, 0}}{m_{\mathrm{default},0} - m_{\mathrm{default}, -\sigma}} 
	\end{align*}
	where $m_{\mathrm{low},0}$ and $m_{\mathrm{high},0}$ are respectively the minimum and maximum median between the default and high-mass (or high-eccentricity) posteriors, $m_{\mathrm{default},0}$ the median of the default posterior, and $m_{\mathrm{default},\pm\sigma}$ respectively the 0.16 and 0.84 quantiles of the default posterior. The authors proposed an arbitrary threshold set at $\Delta_{M_{\mathrm{degen}}} = 1.3$, below which they deemed masses to be well constrained.
	
	\begin{table}[!h]
		\centering
		\setlength{\tabcolsep}{10pt} 
		\renewcommand{\arraystretch}{1.5} 
		\caption{Priors used for the photodynamical fits.}
		\label{tab:priors}
		\begin{tabular}{lll}
			\toprule
			Prior &Mass  & Eccentricity \\
			\midrule
			Default &   Log-uniform  &  $\beta$-distribution\\ 
			High eccentricity  & Log-uniform                                & Uniform \\ 
			High mass &  Uniform                           & Log-uniform\\ 
			\bottomrule 
		\end{tabular}
		\tablefoot{Prior distributions of the planetary masses and eccentricities. The $\beta$-distribution follows \citet{kipping_parametrizing_2013}.}
	\end{table}
	
	\subsection{Resonant state}
	In the formalism of the second fundamental model of resonance \citep{henrard_second_1983}, it is possible to reduce the Hamiltonian of a two-planet system in first-order resonance, when nearly coplanar and of low eccentricity, to a one-degree-of-freedom model. Using the development of \cite{leleu_submitted}, the Hamiltonian then reads:
	\begin{equation}
		\label{eq:Hamiltonian}
		H = 2X - \frac{1}{4}\left(X^2+Y^2 - 3(\delta+1) \right)^2
	\end{equation} 
	For $\delta < 0$, the system possesses a single equilibrium point and no separatrix. There is a bifurcation located at $\delta = 0$, giving two separated equilibrium points for $\delta > 0$, one being encompassed within a separatrix. In this formalism, the phase space is resonant if $\delta > 0$ and the system sits within the separatrix. Slow evolution usually brings systems into resonance during the system formation due to the planet-disk interaction via migration, following the existing resonant fixed points into the separatrix, indicated with the black line in Figure \ref{fig:resonantstate} \citep{goldberg_dynamics_2023}. \cite{ataiee_pushing_2021} showed that overstability can occur during the migration phase, throwing the system out of resonance, but the timescale of these events is much shorter than the disk lifetime, hence it is likely that systems undergoing such processes end up back in resonance.  Conversely, for $\delta < 0$, slow dissipative mechanisms can bring a system out of resonance following the resonant fixed points of the separatrix. Tidal dissipation can bring the system out of strict resonance by dissipation in the inner planet \citep{lithwick_resonant_2012, batygin_dissipative_2012, delisle_dissipation_2012, delisle_tidal_2014}. Systems far from equilibrium points usually originate from a stronger disruptive event in their formation history. Such events can involve dynamical instability and collisions \citep{izidoro_breaking_2017, goldberg_architectures_2022} or planetesimal scattering \citep{chatterjee_planetesimal_2015}. Hence, the resonant state of a system is indeed a convenient tool to infer some information about the formation history of the system \citep{leleu_submitted}.

	\subsection{Kepler-404b and Kepler-404c}
	\begin{figure}[!h]
		\centering
		\includegraphics[width=0.99\linewidth]{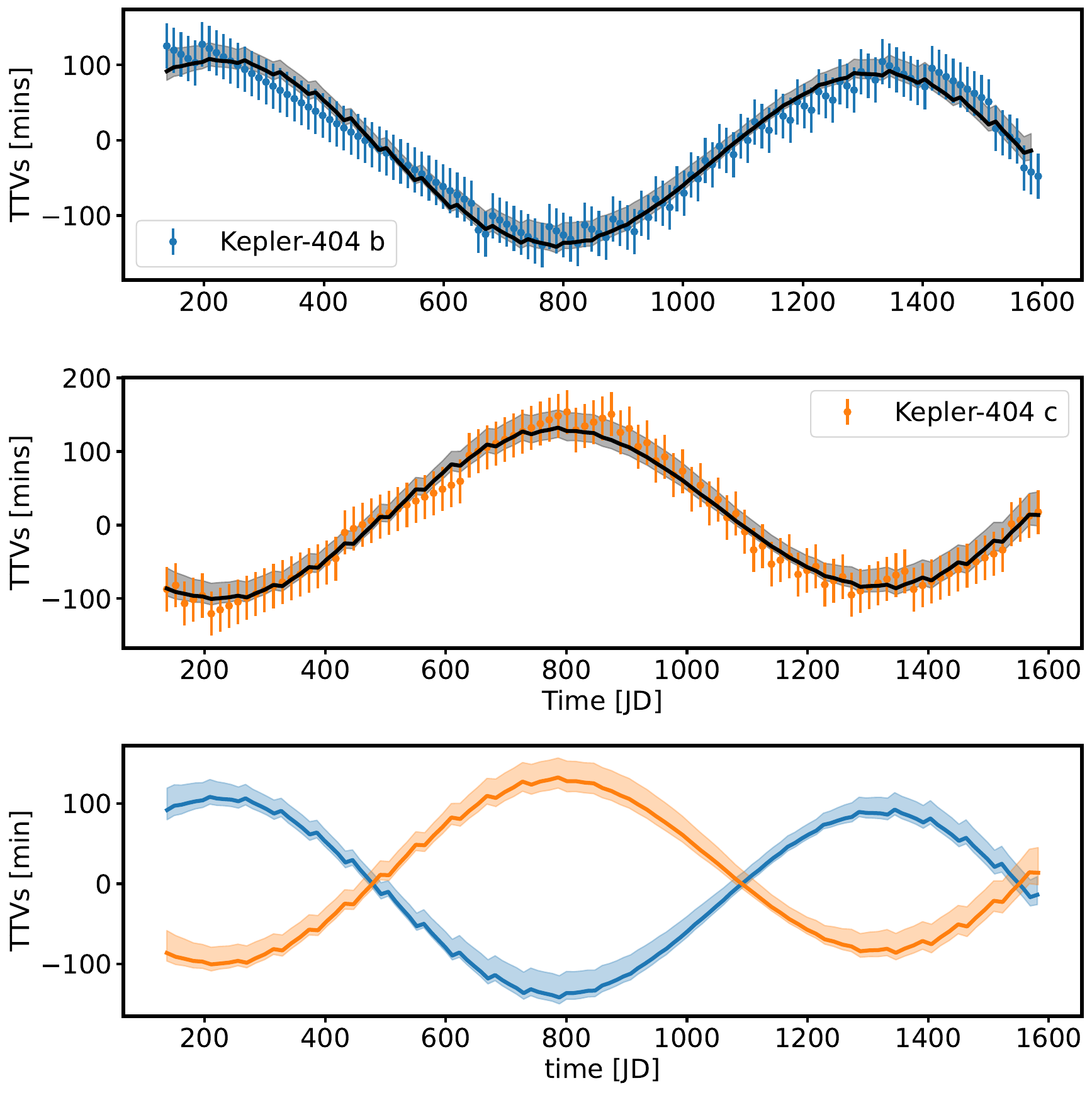}
		\caption{
			O-C plot of the RIVERS.radon proxy transit times in colour and corresponding photodynamical fit timings in black and grey, shown respectively for Kepler-404b and Kepler-404c. The bottom plot shows both fits superimposed, clearly displaying the opposition of phase of the TTVs of both planets.
		}
		\label{fig:ttv_fit}
	\end{figure}
	
	\begin{figure}[!h]
		\centering
		\includegraphics[width=0.9\linewidth]{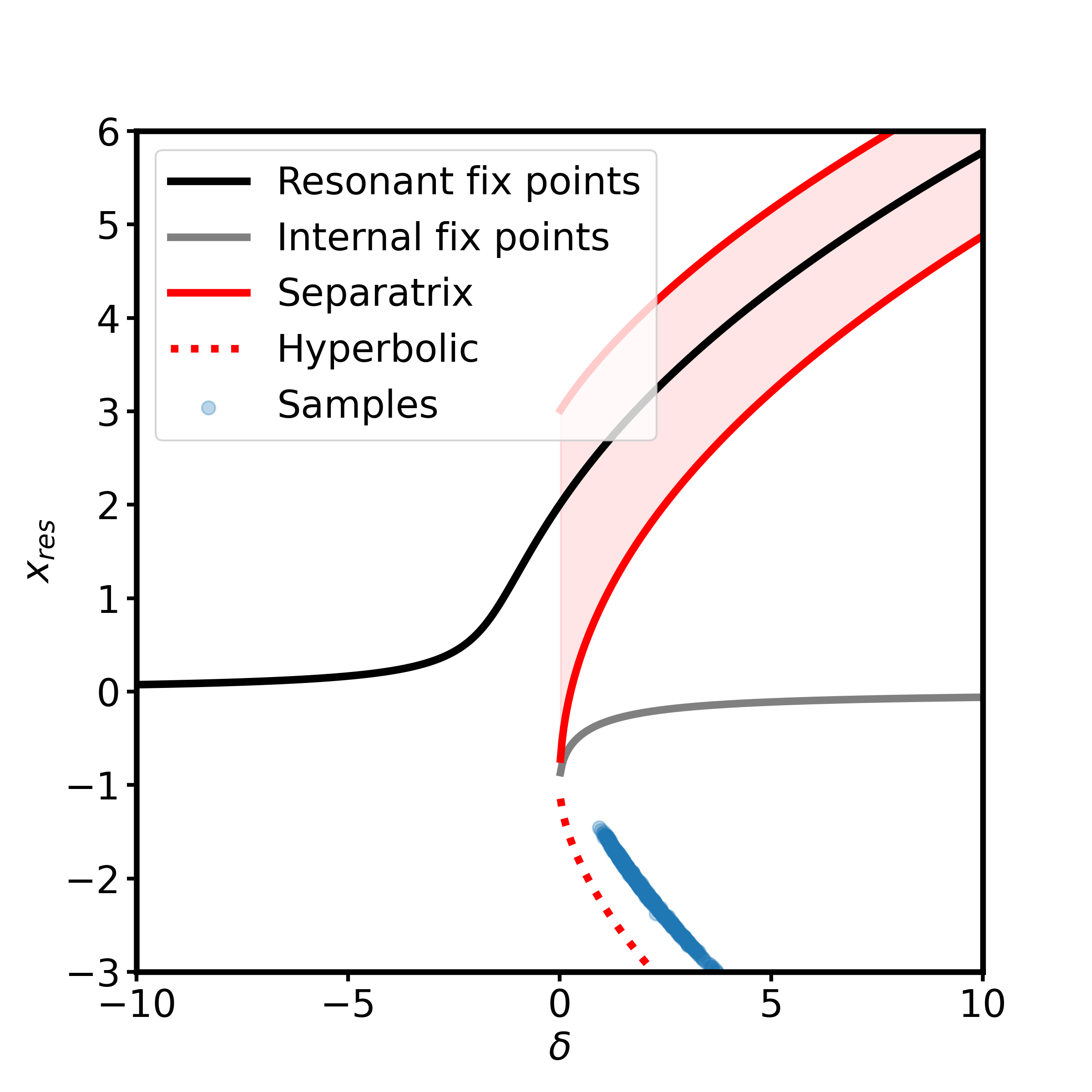}
		\caption{Section of the $(X_{\mathrm{res}}, \delta)$ plane of the Hamiltonian (Eq. \ref{eq:Hamiltonian}) with the separatrix, equilibrium points, and the position of the posteriors of the Kepler-404b and Kepler-404c pair of planets.}
		\label{fig:resonantstate}
	\end{figure}

	Following the fitting process presented in Section \ref{sect:analysis}, we retrieve the transit times with a photodynamical fit.
	Figure \ref{fig:ttv_fit} displays in coloured points the proxy timings retrieved with the RIVERS.radon method, with the 30 min uncertainties inherited from the river diagrams. The black lines show the median of the photodynamical timings and the grey shaded area is the $1\sigma$ deviation of the fit.
	The bottom plot of Figure \ref{fig:ttv_fit} shows the superimposed TTVs of both planets. The TTVs are in clear opposition of phase, with amplitudes of $\mathrm{TTV}_{\mathrm{A,b}}=117.64$ min and $\mathrm{TTV}_{\mathrm{A,c}}=110.09$ min, and periodicities of $\mathrm{TTV}_{\mathrm{P,b}}=1095.5$ days and $\mathrm{TTV}_{\mathrm{P,c}}=1098.9$ days, respectively for Kepler-404b and Kepler-404c. The sawtooth shape of the fit shows a slight chopping effect, caused by the conjunctions of the planets. This effect is valuable for better constraining the planetary masses. However, in this case, it is unclear whether the chopping here is constrained by the data or emerges from the modelling.

\begin{figure}[!h]
		\centering
		\includegraphics[width=0.75\linewidth]{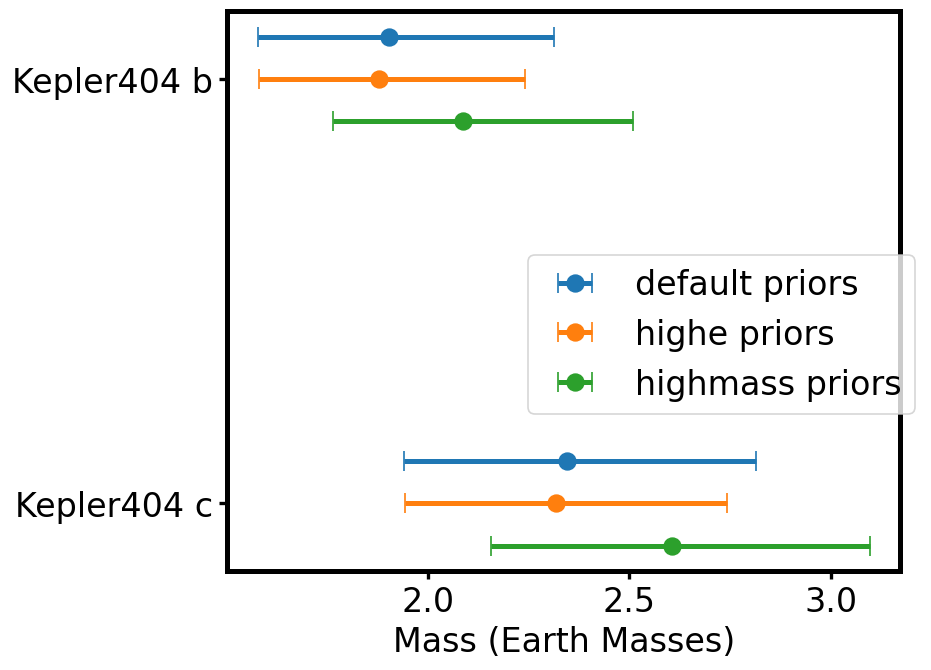}
		\caption{Median and $1\sigma$ quantiles of the posterior distribution of the masses of each planet for the different priors.}
		\label{fig:posteriors}
	\end{figure}

	Table \ref{tab:posteriors} shows the median and $1\sigma$ results of the photodynamical fit samples, with each column corresponding to one of the priors presented in Section \ref{sect:priors}. Figure \ref{fig:posteriors} shows the posterior distribution of the masses of each planet for the different priors. The default and high-eccentricity priors agree closely with each other, while the high-mass prior yields, as expected, higher masses.
	Using the robustness criterion presented in Section \ref{sect:priors}, we compute the $\Delta_M$ for both planets for both priors, and we get $\Delta_{M_\mathrm{b}}=0.08$ and $\Delta_{M_\mathrm{c}}=0.07$ for the high-eccentricity prior and $\Delta_{M_\mathrm{b}}=0.45$ and $\Delta_{M_\mathrm{c}}=0.56$ for the high-mass prior. As all values are below the threshold of $\Delta_M = 1.3$ for both priors, we conclude that the masses are robust to the mass-eccentricity degeneracy.
	
	The photodynamical fit converges to a pair of comparable planets, with similar masses, respectively for Kepler-404b and Kepler-404c  of $1.91_{-0.33}^{+0.41}$ and $2.34_{-0.40}^{+0.47} M_{\oplus}$, and very similar radii of $1.68_{-0.09}^{+0.09}$ and $1.69_{-0.10}^{+0.10} R_{\oplus}$. These similarities in masses and radii suggest that this system has an architecture known as ``peas in a pod'' \citep{weiss_california-kepler_2018}, where planets within the same system are similar in mass and radius, and usually spatially compact. This configuration has been observed to be even more pronounced in near-resonant systems \citep{goyal_enhanced_2023}.
	\begin{figure}[!h]
		\centering
		\includegraphics[width=0.99\linewidth]{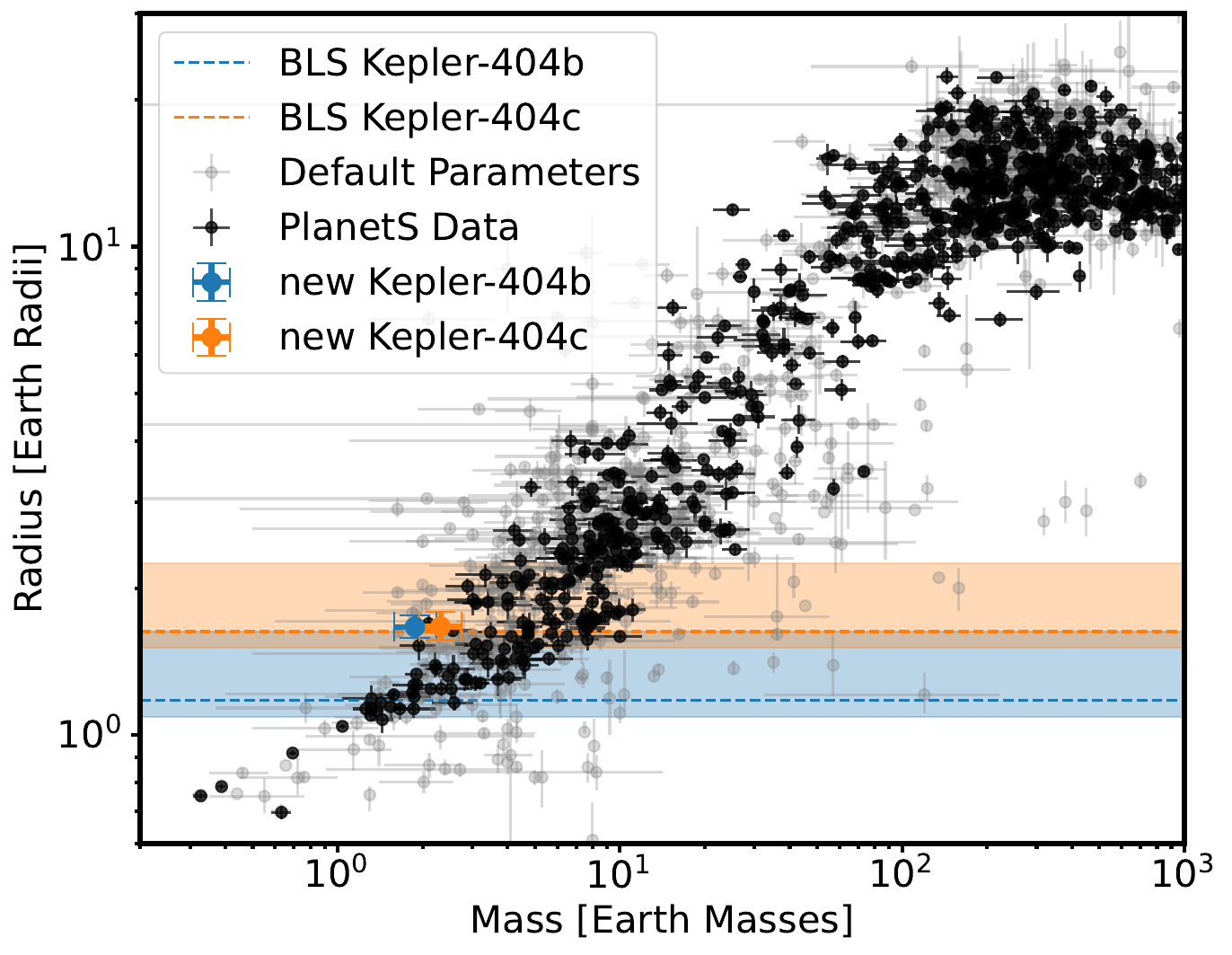}
		\caption{
			Mass-radius plot of currently known exoplanets at the time of writing. The grey points come from the NASA Exoplanet Archive, the black points are from the PlanetS catalogue 
			and the coloured areas and points show the archive and new characterisations of the pair of planets in the Kepler-404 system.
		}
		\label{fig:mass_radius}
	\end{figure}

	Figure \ref{fig:mass_radius} shows the mass-radius diagram of the currently known exoplanet population. The grey points are taken from the NASA Exoplanet Archive, and the black points from the PlanetS catalogue \citep{parc_super-earths_2024}. The previous radii and uncertainties from the archive solution are shown with the coloured area, while the masses and radii from the fit are visible in coloured scatter points. Both Kepler-404b and Kepler-404c sit on the upper edge of the bulk population, indicating that these planets seem to be less dense than their counterparts. This is in agreement with the work of \cite{leleu_resonant_2024}, which showed that resonant systems tend to be statistically less dense than non-resonant systems due to different formation pathways and histories. 
	
	The instantaneous periods are respectively $P= 11.83707_{-0.00058}^{+0.00046}$ days and  $P= 14.74291_{-0.00068}^{+0.00069}$ days. We can estimate the mean period of each planet by integrating each sample forward for $20\,000$ days with TTVFast \citep{deck_ttvfast_2014} to get the expected transit times from the given configuration of the sample, fitting these transit times with a linear ephemeris and measuring the slope. We obtain for Kepler-404b $P_{\mathrm{mean}}  = 11.83243_{-0.000093}^{+0.000083}$ days and for Kepler-404c $P_{\mathrm{mean}}=14.74791_{-0.00012}^{+0.00013}$ days. The ratio of the mean periods is $P_{\mathrm{r}} = 1.24639_{-0.000016}^{+0.000017}$, which is just below the 5:4 commensurability of  $1.25$. Near-resonant systems with a period ratio below the resonant one are not expected to originate from slow dissipative processes, such as tides. Using the mean periods and the resonance $i:j$, we can compute the super-period and find it to be close to the periodicity of the TTVs themselves, as expected for a near-resonant system.
	\begin{equation}
		P_{\mathrm{TTV}} = \frac{1}{\left|\frac{i}{P_{\mathrm{out}}}-\frac{j}{P_{\mathrm{in}}}\right|} = 1023.46_{-4.39}^{+4.82} \text{ days}
	\end{equation}
	The inclinations of both planets are respectively $I_\mathrm{b}=88.75_{-0.18}^{+0.23}$ and $I_\mathrm{c}=88.62_{-0.13}^{+0.15}$ deg, with a minimal mutual inclination of $\Delta I=0.15_{-0.11}^{+0.19}$ deg. Low mutual inclinations (i.e. nearly coplanar orbits) are another feature expected for resonant pairs of planets \citep{agol_refining_2021, leleu_resonant_2024}. 
	
	Figure \ref{fig:resonantstate} shows where each sample of the Kepler-404 pair of planets sits on the $(X_{\mathrm{res}}, \delta)$  plane after computing the transformation leading to Equation \ref{eq:Hamiltonian}. The samples sit in the region of the phase space with $\delta > 0$, next to the hyperbolic fixed point, close to the internal fixed point. Hence, the system is formally not in resonance as not a single sample is located within the separatrix. 
	
	\subsection{Smearing effect}
	The archive solutions were obtained by folding the light curve on a linear ephemeris, thus they are affected by the TTV smearing. We can quantify that by comparing the archive values to our photodynamical fit, visible in Table \ref{tab:radius_comparison}.
	\begin{table*}
			\centering
			\caption{Transit parameters of Kepler-404b and Kepler-404c from the archive and from this work.}
			\label{tab:radius_comparison}
		\begin{tabular}{lcccccc}
			\toprule
			& \multicolumn{3}{c}{Kepler-404b} & \multicolumn{3}{c}{Kepler-404c} \\
			\cmidrule(lr){2-4} \cmidrule(lr){5-7}
			Parameter & Archive & This work & Deviation & Archive & This work & Deviation \\
			\midrule
			$R_\mathrm{p}/R_\star$      & $0.01261_{-0.00087}^{+0.00099}$ & $0.01906_{-0.00083}^{+0.00070}$ & $6.5\,\sigma$
			& $0.01740_{-0.00411}^{+0.00543}$ & $0.01908_{-0.00093}^{+0.0010}$ & $0.3\,\sigma$ \\[2pt]
			$b$                & $0.266_{-0.266}^{+0.201}$ & $0.563_{-0.094}^{+0.069}$ & $1.5\,\sigma$
			& $0.876_{-0.584}^{+0.004}$ & $0.723_{-0.065}^{+0.055}$ & $0.3\,\sigma$ \\[2pt]
			$R_\mathrm{p}$ [$R_\oplus$] & $1.18_{-0.09}^{+0.45}$ & $1.68_{-0.09}^{+0.09}$ & $1.1\,\sigma$
			& $1.63_{-0.12}^{+0.62}$ & $1.69_{-0.10}^{+0.10}$ & $0.1\,\sigma$ \\
			\bottomrule
		\end{tabular}
		\tablefoot{Archive values are from the NASA Exoplanet Archive solution, obtained from a linear-ephemeris fold. This work gives the median and $\pm1\sigma$ of the default photodynamical fit.}
		\end{table*}
	
	For Kepler-404b, Table \ref{tab:radius_comparison} shows that the radius ratio changes from $R_\mathrm{p}/R_\star = 0.01261_{-0.00087}^{+0.00099}$ in the archive to $0.01906_{-0.00083}^{+0.00070}$ in our fit, a significant revision of $6.5\sigma$, which is likely driven by the correction of the recovered transit depth when stacking the transits on the individual transit times, which removes the dilution caused by the smearing. For Kepler-404c, the radius ratio goes from $0.01740_{-0.00411}^{+0.00543}$ to $0.01908_{-0.00093}^{+0.0010}$, the two values are consistent within $0.3\sigma$, the estimation of the transit depth of the planet was somewhat less affected by the smearing. The impact parameters are $b=0.266_{-0.266}^{+0.201}$ from the archive against $b=0.563_{-0.094}^{+0.069}$ in our fit for Kepler-404b, this is a $1.5\sigma$ change. For Kepler-404c, the value goes from $b=0.876_{-0.584}^{+0.004}$ to $b=0.723_{-0.065}^{+0.055}$, a $0.3\sigma$ difference. As these values lie away from the grazing regime, they are unlikely to affect the estimation of the radius ratio and planetary radii.
	Finally, the planetary radius changes from $1.18_{-0.09}^{+0.45}\,R_{\oplus}$ to $1.68_{-0.09}^{+0.09}\,R_{\oplus}$ for Kepler-404b, a $1.1\sigma$ change, and from $1.63_{-0.12}^{+0.62}\,R_{\oplus}$ to $1.69_{-0.10}^{+0.10}\,R_{\oplus}$ for Kepler-404c, a $0.1\sigma$ change. The small deviations are mainly due to the large upper uncertainty of the stellar radius, which propagates into the planetary radius in the archive solution. In addition, the stellar radius adopted in this work ($0.813\,R_{\odot}$) is smaller than the archive one ($0.857\,R_{\odot}$) and therefore scales down the retrieved planetary radius, partly counterbalancing the correction of the smearing. The characterisation of Kepler-404b was thus clearly affected by the smearing effect, as shown by the revision of its radius ratio, whereas for Kepler-404c the archive solution is too poorly constrained for a comparable statement to be made. 
	
	The correction of the smearing effect is also reflected in the stacked S/N that can be computed using Equation \ref{eq:snr_fold}, giving $\mathrm{S/N} = 20.99$ and $\mathrm{S/N} = 14.88$ respectively for Kepler-404b and Kepler-404c. The previously found S/Ns at the time of writing were $\mathrm{S/N} = 9.10$ for the confirmed signal of Kepler-404b,  $\mathrm{S/N} = 12.70$ for the false positive KOI-2768.03, and $\mathrm{S/N} = 8.30$ for Kepler-404c. 
	
	Finally, we stress that the effect of the smearing on the characterisation of a planet based on a linear ephemeris is difficult to predict. The folded transit results from the true transit shape being spread over the residual phase offsets of the individual transits, which depend on the amplitude and periodicity of the TTVs, and on where the linear ephemeris intersects the TTV curve. The resulting distorted profile is then fitted by a transit model that will be affected by degeneracies between depth, duration, impact parameter and stellar density, so the bias does not systematically translate into a shallower transit, smaller radius or longer duration. The archive solutions of the Kepler-404 system at the time of writing illustrate this behaviour well: the estimated transit duration of KOI-2768.03 is $6.15$ h, almost twice that of Kepler-404b ($3.34$ h), although the two ephemerides differ by only $4$ min in period and capture a signal belonging to the same planet, while the transit duration of Kepler-404c is $1.25$ h, roughly half the duration our fit yields (calculated around $2.7$ h).

	\section{Discussion and conclusion}
	In this work, we highlight the detection bias caused by TTVs when using phase-folding algorithms on planets with low S/N. We present how the 2D representations known as river diagrams make it possible to display the planetary transit signal. With the case of Kepler-404, we show that the BLS solutions of Kepler-404b and Kepler-404c are indeed affected by the TTVs caused by their near-resonant configuration.
	We introduce the RIVERS.radon method, an algorithm that relies on river diagrams to extract the transit tracks from the image with the use of 2D cross-correlation. Applying this novel method to the river diagram of both planets, we successfully extract the approximate transit times, taking into account the TTVs of the planets. With the new individual proxy timings used to get initial conditions on the orbital parameters of the system, we characterise both planets using a photodynamical fit, correcting the bias on the radius of the planets due to the smearing effect of TTVs, and retrieve a complete characterisation of the system, including the planetary masses.
	The proximity of the ephemerides of Kepler-404b and KOI-2768.03, whose orbital periods differ by only $\sim$4 minutes, was a strong hint that these two signals were connected. A visual inspection of the river diagram of Kepler-404b in Figure \ref{fig:rd_kepler-404b_and_c} showed that both ephemerides intersect on the same transit track, each matching a different segment. The BLS solutions picked two local approximations of one quasi-periodic signal. The transit times resulting from the photodynamical fit closely follow the proxy timings of RIVERS.radon over the whole baseline with no apparent divergence, as shown in Figure \ref{fig:ttv_fit}. Additionally, a subsequent BLS search of the residual light curve, after removing the photodynamical model of both planets, yields no significant signal at or near the KOI-2768.03 ephemeris. This confirms that the signal composed of the original Kepler-404b and KOI-2768.03 ephemerides is successfully explained by a single planet. Finally, stacking the transits on the photodynamical timings yields an S/N of 20.99, exceeding both original signals, respectively 9.10 and 12.70 for Kepler-404b and KOI-2768.03, as expected once the smearing effect on the stacking of the transits caused by the TTVs is corrected. The fits converge to a pair of similar planets, respectively for Kepler-404b and Kepler-404c, of radii $1.68_{-0.09}^{+0.09}$ and $1.69_{-0.10}^{+0.10} R_{\oplus}$ and of masses $1.91_{-0.33}^{+0.41}$ and $2.34_{-0.40}^{+0.47} M_{\oplus}$. We show that according to the robustness criterion presented in Section \ref{sect:priors}, our masses are robust to the mass-eccentricity degeneracy, and that the radius of Kepler-404b was underestimated due to the smearing effect caused by the TTVs. These similarities in masses and radii suggest that the system is of a type known as ``peas in a pod''. With the new individual transit times, we retrieve TTVs in clear opposition of phase, with amplitude and period ($\mathrm{TTV}_\mathrm{A}$, $\mathrm{TTV}_\mathrm{P}$)  of ($117.64$ min, $1095.5$ days) and ($110.09$ min, $1098.9$ days) for Kepler-404b and Kepler-404c, respectively. As expected for near-resonant planets, the super-period, $P_{\mathrm{TTV}} = 1023.46_{-4.39}^{+4.82} \text{ days}$, is relatively close to the periodicity of the TTVs of the planets. Using these transit times, we compute new S/N values of 20.99 and 14.88. These S/N values are high enough for the planets to be detected despite the TTVs, yet the TTVs still caused the false-positive detection.
	
	This system exhibits characteristics consistent with near-resonant systems. The low density of the planets is in agreement with the observed trend that resonant systems are often less dense than the bulk population \citep{leleu_resonant_2024}. The mutual inclination of the planets is low, which is also a trend observed in near-resonant pairs of planets \citep{agol_refining_2021, leleu_resonant_2024}. Although the period ratio, $P_{\mathrm{r}} = 1.24639_{-0.000016}^{+0.000017}$, is very close to the 5:4 commensurability of $1.25$, the system is actually non-resonant. Slow physical processes usually bring systems out of resonance by reducing the eccentricities, e.g. tides \citep{ lithwick_resonant_2012, batygin_dissipative_2012, delisle_dissipation_2012, delisle_tidal_2014}. This feature indicates that the system may have undergone post-disk instability. 
	
	Figure \ref{fig:resonantstate} shows that the system is formally non-resonant, as all samples sit outside the separatrix, with $\delta >0$. Typically, systems are either in resonance or at low eccentricities, but this system is observed out of resonance and relatively far away from the fixed points. Overstabilities are capable of breaking resonant chains during the migration phase of the formation of the system \citep{goldreich_overstable_2014, delisle_stability_2015}. However, as they occur during the migration phase, the system is brought back into resonance on a timescale usually shorter than the disk lifetime \citep{ataiee_pushing_2021, deck_migration_2015}. Thus, while possible, it is unlikely that the overstability occurred right as the disk dissipated. Post-disk instabilities could explain the relatively large distance from the fixed points of the system. However, this idea is somewhat at odds with the observed coplanarity of the system, as planet-planet scattering tends to increase the mutual inclination of the planets \citep{leleu_resonant_2024}. 
	
	The RIVERS.radon method is complementary to other methods with the same aim, namely the QATS algorithm introduced in \cite{carter_quasiperiodic_2013} and the RIVERS.deep approach from \cite{leleu_alleviating_2021}. Both RIVERS approaches are used as first estimators for the initial conditions of the photodynamical fit and require an additional step to translate the solutions from the river-diagram space to the time domain. The RIVERS.radon method finds its strength in the use of a set of local periodic solutions to approximate the transit times, rendering the method capable of finding planetary signals with significant TTVs while reducing the risk of a genuine signal being confused with the background noise, at the cost of a non-negligible number of hyper-parameters to be fine-tuned to find the optimal solutions. On the other hand, RIVERS.deep is able to recover individual transits down to lower S/N, at the cost of interpretability, as it is a deep-learning model. Finally, QATS is more versatile, as it provides an automatic optimal stacked transit sequence optimising the S/N for a chosen range of parameters from the time series directly, at the cost of a greater sensitivity to the background noise as the algorithm retrieves candidates freely in a given tolerance range. Hence, these methods are complementary to one another and can be used alone or collectively according to the task, from a blind planet detection search to fine-tuning the individual transit times of a known planet.
	
	Finally, this work shows the complexity involved in characterising (near)-resonant systems. Considerable efforts have been put towards improving our understanding of the formation and evolution of these systems through theoretical work. However, the constraints on theoretical models can be limited by the number of robustly characterised (near)-resonant systems available for comparison. The system of Kepler-404 is a good example of how such systems can be mischaracterised or simply missed due to the observational bias caused by the TTVs. Hence, a systematic characterisation of these systems is needed. Additionally, re-analysing transit-survey data, notably from Kepler and the TESS continuous viewing zones, is necessary to alleviate the TTV bias and to prepare for the upcoming PLATO mission.

	\begin{acknowledgements}
		This work has been carried out within the framework of the National Centre of Competence in Research PlanetS supported by the Swiss National Science Foundation and benefited from the seed-funding programme of the Technology Platform of PlanetS. The authors acknowledge the financial support of the SNSF. 
		Y.E., M.R., A.L. acknowledge support of the Swiss National Science Foundation under grant number TMSGI2\_211697.
	\end{acknowledgements}
	\bibliographystyle{aa}
	\bibliography{citations_aa} 
	
	\begin{appendix}
	\section{Estimation of the transit times}
	\label{sect:appendix}

	The confidence map $R \in \mathbb{R}^{H \times W}$ has the same dimensions as the input river diagram. It is a direct per-pixel confidence of a transit location. Because the river diagram can be created with flux but also with time, we also have a direct correspondence of time for each confidence map element.  Let $T^R \in \mathbb{R}^{H \times W}$ be the time map, where each pixel $T^R(p,q)$ is the corresponding time value of the matching pixel $(p,q)$ of the river diagram, thus by extension the time value of the matching pixel of the confidence map $R$. We define the continuous confidence function $\phi(t)$ as:
	\begin{equation}
		\phi(t)
		=
		\sum_{p=0}^{H-1} \sum_{q=0}^{W-1}
		R(p,q)\,
		\delta\big(t - T^R(p,q)\big),
	\end{equation}
	In practice, the time series is discrete, with a sequence of times $\{t_k\}$ of cadence smaller than the river diagram bin size. Thus the confidence function is approximated as:
	\begin{equation}
		\phi(t_k)
		=
		R(p^*, q^*) \quad \text{with} \quad (p^*, q^*) = \arg\min_{(p,q)} \left|T^R(p,q) - t_k\right|.
	\end{equation}
	We extract the best candidate transit time $t_0$ with the highest confidence 
	\begin{equation}
		\widehat{t}_0
		=
		\arg \max_{t_k} \phi(t_k).
	\end{equation}
	Starting from $t_0$ we look backward and forward in the confidence time series to extend the transit-time sequence across successive periods within a given tolerance. The adjacent interval, for a given transit time $t_i$, is given by
	\begin{equation}
		\mathcal{I}_{i\pm1} = [t_i \pm P(1- \epsilon), t_i \pm P(1 +\epsilon)]  
	\end{equation}
	where $\epsilon$ is a tolerance accounting for the effect of TTVs. The corresponding best candidate transit time $t_i$ of the given interval $\mathcal{I}_i$ is found with
	\begin{equation}
		\label{eq:candidate}
		\widehat{t}_{i \pm 1}
		=
		\arg \max_{t_k \in \mathcal{I}_{i \pm 1}} \phi(t_k)
	\end{equation} 
	If no direct adjacent candidate is found, the search is extended to multiple periods ahead, until a new candidate is found or the start or end of the time series is reached. Finally, we end up with the transit-time sequence $\{t_i\}_{i=1}^{M}$, with $M$ the number of transits found. 
	\section{Photodynamical fit posteriors} 
		\begin{table*}
		\centering
		\renewcommand{\arraystretch}{1.08}
		\caption{Posteriors of the photodynamical fits of Kepler-404.}
		\label{tab:posteriors}
		\begin{tabular}{llll}
			\toprule
			\multicolumn{4}{c}{Kepler-404b} \\
			\midrule
			Parameter & Default & High eccentricity & High mass \\
			\midrule
			\multicolumn{4}{l}{\textit{Fitted parameters}} \\
			$\lambda$ [deg] & $44.82_{-0.99}^{+0.98}$ & $44.63_{-9.29}^{+4.83}$ & $44.78_{-0.63}^{+0.59}$ \\
			$P$ [days] & $11.83707_{-0.00058}^{+0.00046}$ & $11.83695_{-0.00060}^{+0.00056}$ & $11.83686_{-0.00057}^{+0.00048}$ \\
			$e\cos \bar{\omega}$ [-1,1] & $0.0046_{-0.0083}^{+0.0078}$ & $0.006_{-0.043}^{+0.089}$ & $0.0046_{-0.0048}^{+0.0043}$ \\
			$e\sin \bar{\omega}$ [-1,1] & $0.009_{-0.011}^{+0.010}$ & $0.03_{-0.03}^{+0.12}$ & $0.0071_{-0.0073}^{+0.0069}$ \\
			$M_\mathrm{p}$ [$10^{-6}\,M_{\odot}$] & $5.7_{-0.98}^{+1.2}$ & $5.6_{-0.89}^{+1.1}$ & $6.3_{-0.97}^{+1.3}$ \\
			$R_\mathrm{p}$ [$R_{\star}$] & $0.01906_{-0.00083}^{+0.00070}$ & $0.01881_{-0.00077}^{+0.00083}$ & $0.01909_{-0.00075}^{+0.00076}$ \\
			$I$ [deg] & $88.75_{-0.18}^{+0.23}$ & $88.87_{-0.27}^{+0.61}$ & $88.76_{-0.17}^{+0.20}$ \\
			\multicolumn{4}{l}{\textit{Derived parameters}} \\
			$a$ [au] & $0.0976_{-0.0046}^{+0.0050}$ & $0.0975_{-0.0048}^{+0.0043}$ & $0.0993_{-0.0048}^{+0.0049}$ \\
			$e$ & $0.015_{-0.011}^{+0.016}$ & $0.06_{-0.05}^{+0.15}$ & $0.011_{-0.010}^{+0.005}$ \\
			$\bar{\omega}$ [deg] & $55.4_{-97.0}^{+42.6}$ & $60.9_{-75.4}^{+56.8}$ & $54.2_{-86.6}^{+22.4}$ \\
			$b$ & $0.563_{-0.094}^{+0.069}$ & $0.51_{-0.27}^{+0.10}$ & $0.563_{-0.084}^{+0.061}$ \\
			$a/R_{\star}$ & $25.87_{-0.97}^{+0.97}$ & $25.80_{-0.94}^{+0.92}$ & $26.09_{-0.97}^{+0.85}$ \\
			$M_\mathrm{p}$ [$M_{\oplus}$] & $1.90_{-0.33}^{+0.41}$ & $1.88_{-0.30}^{+0.36}$ & $2.09_{-0.32}^{+0.42}$ \\
			$R_\mathrm{p}$ [$R_{\oplus}$] & $1.683_{-0.091}^{+0.090}$ & $1.666_{-0.083}^{+0.090}$ & $1.708_{-0.097}^{+0.091}$ \\
			$\rho_\mathrm{p}$ [$\rho_{\oplus}$] & $0.397_{-0.069}^{+0.098}$ & $0.410_{-0.080}^{+0.084}$ & $0.421_{-0.067}^{+0.090}$ \\
			$T_0$ [BJD$-$2454833] & $137.6518_{-0.0075}^{+0.0078}$ & $137.6528_{-0.0077}^{+0.0070}$ & $137.6541_{-0.0075}^{+0.0068}$ \\
			S/N &  20.99 &  \ldots&\ldots  \\
			\midrule
			\multicolumn{4}{c}{Kepler-404c} \\
			\midrule
			Parameter & Default & High eccentricity & High mass \\
			\midrule
			\multicolumn{4}{l}{\textit{Fitted parameters}} \\
			$\lambda$ [deg] & $65.69_{-0.79}^{+0.93}$ & $65.54_{-8.47}^{+4.15}$ & $65.74_{-0.46}^{+0.64}$ \\
			$P$ [days] & $14.74291_{-0.00068}^{+0.00069}$ & $14.74295_{-0.00076}^{+0.00076}$ & $14.74321_{-0.00073}^{+0.00071}$ \\
			$e\cos \bar{\omega}$ [-1,1] & $-0.0029_{-0.0075}^{+0.0070}$ & $-0.001_{-0.037}^{+0.078}$ & $-0.0030_{-0.0048}^{+0.0033}$ \\
			$e\sin \bar{\omega}$ [-1,1] & $-0.004_{-0.011}^{+0.008}$ & $0.01_{-0.02}^{+0.11}$ & $-0.0043_{-0.0079}^{+0.0044}$ \\
			$M_\mathrm{p}$ [$10^{-6}\,M_{\odot}$] & $7.0_{-1.2}^{+1.4}$ & $7.0_{-1.1}^{+1.3}$ & $7.8_{-1.3}^{+1.5}$ \\
			$R_\mathrm{p}$ [$R_{\star}$] & $0.01908_{-0.00093}^{+0.0010}$ & $0.0189_{-0.0011}^{+0.0011}$ & $0.01907_{-0.00094}^{+0.00086}$ \\
			$I$ [deg] & $88.62_{-0.13}^{+0.15}$ & $88.63_{-0.16}^{+0.18}$ & $88.65_{-0.15}^{+0.14}$ \\
			\multicolumn{4}{l}{\textit{Derived parameters}} \\
			$a$ [au] & $0.1130_{-0.0053}^{+0.0058}$ & $0.1129_{-0.0055}^{+0.0050}$ & $0.1149_{-0.0056}^{+0.0057}$ \\
			$e$ & $0.012_{-0.010}^{+0.012}$ & $0.05_{-0.04}^{+0.12}$ & $0.0080_{-0.0076}^{+0.0063}$ \\
			$\bar{\omega}$ [deg] & $-87_{-44}^{+187}$ & $39_{-157}^{+88}$ & $-115_{-19}^{+180}$ \\
			$b$ & $0.723_{-0.065}^{+0.055}$ & $0.711_{-0.078}^{+0.070}$ & $0.711_{-0.066}^{+0.059}$ \\
			$a/R_{\star}$ & $29.95_{-1.12}^{+1.12}$ & $29.87_{-1.09}^{+1.05}$ & $30.20_{-1.13}^{+0.98}$ \\
			$M_\mathrm{p}$ [$M_{\oplus}$] & $2.34_{-0.40}^{+0.47}$ & $2.32_{-0.37}^{+0.42}$ & $2.60_{-0.45}^{+0.49}$ \\
			$R_\mathrm{p}$ [$R_{\oplus}$] & $1.69_{-0.10}^{+0.10}$ & $1.67_{-0.10}^{+0.12}$ & $1.71_{-0.10}^{+0.10}$ \\
			$\rho_\mathrm{p}$ [$\rho_{\oplus}$] & $0.48_{-0.09}^{+0.12}$ & $0.49_{-0.10}^{+0.13}$ & $0.52_{-0.10}^{+0.13}$ \\
			$T_0$ [BJD$-$2454833] & $137.1912_{-0.0075}^{+0.0088}$ & $137.1914_{-0.0081}^{+0.0081}$ & $137.1890_{-0.0070}^{+0.0089}$ \\
			S/N & 14.88 & \ldots &\ldots  \\
			\midrule
			\multicolumn{4}{c}{Kepler-404} \\
			\midrule
			Parameter & Default & High eccentricity & High mass \\
			\midrule
			\multicolumn{4}{l}{\textit{Fitted parameters}} \\
			$\rho_{\star}$ [$\rho_{\odot}$] & $1.66_{-0.18}^{+0.19}$ & $1.64_{-0.17}^{+0.18}$ & $1.70_{-0.18}^{+0.17}$ \\
			$R_{\star}$ [$R_{\odot}$] & $0.813_{-0.030}^{+0.028}$ & $0.811_{-0.026}^{+0.030}$ & $0.820_{-0.028}^{+0.031}$ \\
			jitter [$10^{-4}$] & $1.7_{-0.11}^{+0.11}$ & $1.7_{-0.11}^{+0.10}$ & $1.7_{-0.10}^{+0.10}$ \\
			Limb darkening $u_1$ & $0.429_{-0.026}^{+0.027}$ & $0.432_{-0.025}^{+0.024}$ & $0.432_{-0.025}^{+0.026}$ \\
			Limb darkening $u_2$ & $0.253_{-0.029}^{+0.030}$ & $0.251_{-0.028}^{+0.028}$ & $0.250_{-0.029}^{+0.030}$ \\
			\bottomrule
		\end{tabular}
		\tablefoot{Median and $\pm 1\sigma$ percentiles of the photodynamical fit samples for the stellar and
			planetary parameters. Orbital elements are given at the reference epoch
			$t_{\mathrm{ref}} = 136.18281$ BKJD (BJD$_{\mathrm{TDB}} - 2454833$).
			Derived parameters are computed from the fitted parameters of each sample;
			$T_0$ is the first transit after $t_{\mathrm{ref}}$.
			Each column corresponds to a different prior used for the fit, the first being the default one, used throughout the paper.}
	\end{table*}
	
\end{appendix}

\end{document}